\documentclass[aps,english,showpacs,11pt]{revtex4}
\usepackage[latin1]{inputenc}
\usepackage{ucs}
\usepackage{amsmath}
\usepackage{amsfonts}
\usepackage{amssymb}
\usepackage{graphicx}
\usepackage{picinpar}
\usepackage[]{hyperref}

\begin{document}
\title{Galactic cold dark matter as a Bose-Einstein condensate of
  WISPs}

\author{M. O. C. Pires}
\email{marcelo.pires@ufabc.edu.br}
\author{J. C. C. de Souza} \email{jose.souza@ufabc.edu.br}

\affiliation{Centro de Ci\^{e}ncias Naturais e Humanas, Universidade
  Federal do ABC, Rua Santa Ad\'elia 166, 09210-170, Santo Andr\'{e}, SP,
  Brazil}

\begin{abstract}
We propose here the dark matter content of galaxies as a cold bosonic
fluid composed of Weakly Interacting Slim Particles (WISPs),
represented by spin-0 axion-like particles and spin-1 hidden bosons,
thermalized in the Bose-Einstein condensation state and bounded by
their self-gravitational potential. We analyze two zero-momentum
configurations: the polar phases in which spin alignment of two
neighbouring particles is anti-parallel and the ferromagnetic phases
in which every particle spin is aligned in the same direction. Using
the mean field approximation we derive the Gross-Pitaevskii equations
for both cases, and, supposing the dark matter to be a polytropic
fluid, we describe the particles density profile as Thomas-Fermi
distributions characterized by the halo radii and in terms of the
scattering lengths and mass of each particle. By comparing this model
with data obtained from 42 spiral galaxies and 19 Low Surface
Brightness (LSB) galaxies, we constrain the dark matter particle mass
to the range $10^{-6}-10^{-4}\; eV$ and we find the lower bound for the
scattering length to be of the order $10^{-14}\; fm$.
\end{abstract}

\pacs{98.80.Cq; 98.80.-k; 95.35.+d}

\maketitle

\section{Introduction}

One of the biggest challenges of modern science is to determine the
constituents of an unusual sort of matter that is at the moment only
observable through its gravitational interaction with usual (barionic)
matter. This so-called dark matter corresponds to about $27\%$ of the
total density energy of the universe at the present time \cite{wmap},
and may dominate the total mass in galaxies.

Amongst the many proposed candidates we can point the WIMPs (Weakly
Interacting Massive Particles)\cite{beringer}. These are particles
that present a very small coupling parameter to barionic matter and at
the same time a very high mass of the order, at least, of GeV. Such a
high mass poses the problem of the stability of these particles in our
Universe.

Another class of dark matter particle candidates, with masses in the
sub-eV range, have been recently proposed as WISPs (Weakly Interacting
Slim Particles) \cite{jaeckel,nel11}. They may include the usual QCD axion,
axion-like particles with a similar coupling of axions but smaller
masses, and low-mass spin-1 bosons.

Axions, hypothetical particles proposed by Peccei and Quinn
\cite{peccei}, are scalar fields which have a nonzero vacuum
expectation value and keep the CP invariance of the strong
interactions in the Lagrangian that originally possesses a $U(1)$
invariance involving all Yukawa couplings. They should have a mass
range of $10^{-6}-1\; eV$. Although the experiments have failed so far
to prove their existence due to the effect of an almost collisionless
scattering with baryonic particles \cite{feng10}, they are an
interesting possibility for cosmology, since low mass axions are
predicted to have been formed shortly after the Big Bang, and may
constitute the dark matter component of the present Universe
\cite{axion01}. Experimental searches are in execution or in the
planning phase \cite{arias, baker}.

Since axions (axion-like particles) are defined as spin-zero bosonic
particles, several authors \cite{sikivie1,boh07, chavanis, chavanis2}
have suggested a cold dark matter fluid composed of a self-gravitating
Bose-Einstein condensate (BEC) of axions (axion-like particles)
constituting the galactic dark matter halo. This fluid is made of
weakly coupled self-interacting particles and presents a huge phase
space density enabling it to suitably describe the density profile of
the galaxy, as long as convenient approximations are assumed. Other
approaches consider noninteracting particles (see, e. g., \cite{lora,
  robles}), with the result that their masses become ultralight ($\sim
10^{-24}\; eV$).

Experiments have showed that even a spinorial gas can reach the BEC
phase \cite{stenger98} and this fact motivates the introduction of a
spin-1 BEC to constitute the dark matter halo. The spin-1 WISPs are
called hidden photons or hidden bosons and their formal derivation is
particularly linked to the string theory framework \cite{jaeckel,
  nel11}. The proposal of these particles as the components of the
dark matter fluid leads to the introduction of additional parameters
to model the halo's density profile, as we shall show latter.
 
Although we do not claim to identify the particle that compose the
dark matter fluid, our treatment allows one to relate the spin-0
particles to axions (axion-like particles) and the spin-1 particles to
hidden photons as depicted in references \cite{jaeckel, nel11}. It is
their spins and their masses that are the relevant features for the
method we implement here.
   
The present paper is organized as follows.  In section \ref{secspin0}
we develop the theory of the spin-0 condensate based on reference
\citep{boh07} and in section \ref{secspin1} we presume the spin-1
condensate in the ferromagnetic and polar phases and, using an
appropriate statistical analysis to the fit of 42 spiral and 19 Low
Surface Brightness (LSB) galaxies, we obtain in section
\ref{secstatistical} the mass and scattering lengths of WISPs. The
discussion and conclusion of our results are in section
\ref{secconclusion}.

\section{Spin-0 particles Bose-Einstein condensate}
\label{secspin0}

In this case, we treat the condensate using a scalar mean field $\phi$,
which has the following Lagrangian density \cite{pethick}:
\begin{eqnarray}
{\cal L}&=&-\frac{\hbar^{2}}{2m}(\nabla\phi^{*})\cdot(\nabla\phi)-
\frac{\hbar}{2i}\left((\phi^{*})(\partial_{t}\phi)-
(\partial_{t}\phi^{*})(\phi)\right)\nonumber
\\ &-&\phi^{*}U\phi-\lambda|\phi|^{4}\; ,
\label{lagrangian}
\end{eqnarray}
where $m$ is the particle mass, $U$ is the external potential produced
by self-gravitational effect, and the term $\lambda|\phi|^{4}$
represents a 2-body point interaction with $\lambda = 2\pi\hbar^2a/m$
proportional to the $s$-wave scattering length $a$.

Using the expression (\ref{lagrangian}) in the Euler-Lagrange equations, we
derive the Gross-Pitaevskii equation for the spin-0 condensate as
\begin{equation}
i\hbar\partial_{t}\phi=-\frac{\hbar^{2}}{2m}\nabla^{2}\phi+U\phi+2\lambda|\phi|^{2}\phi\,.
\label{GPE}
\end{equation}

We regard $|\phi|^2$ as the particle density of the condensate,
$\rho({\bf x})$, thus we normalize it to the particle number $N$
($\int|\phi|^{2}d^{3}x=N$). Assuming that the particle number is
conserved in the system, we can parametrize the wave function by
$\phi(\textbf{x},t) = e^{-i\mu t/\hbar}\sqrt{\rho({\bf
    x})}e^{i\frac{S({\bf x})}{\hbar}}$ where $\mu$ is the chemical
potential and the $S({\bf x})$ is the wave function's quantum phase \citep{boh07}.

The Gross-Pitaevskii equation (\ref{GPE}) splits in two parts, one corresponding
to the imaginary part
\begin{equation}
\nabla\cdot(\rho\vec{v})=0\;,
\end{equation}
with $\vec{v}=\frac{\nabla S}{m}$, and another one corresponding to
the real part of the equation
\begin{equation}
\nabla(2\lambda \rho + \frac{1}{2m}|\nabla S|^2 + W + U)=0\;,
\label{qeuler}
\end{equation}
where
$W=-\frac{\hbar^{2}}{2m}\frac{\nabla^{2}\sqrt{\rho}}{\sqrt{\rho}}$ is
the quantum potential. 

For a self-gravitating Bose-Einstein condensate, the external
potential $U$ obeys the Poisson equation
\begin{equation}
\nabla^2 U= 4\pi G \rho_m\;,
\end{equation}
where $\rho_m = m\rho$ is the condensate mass density. In the case of
a static condensate, for which $\vec{v}\equiv 0$, we can have the
polytropic fluid with equation of state
\begin{equation}
p = K\rho_m^\Gamma = K\rho_m^{1+\frac{1}{n}}\,,
\end{equation}
where $n$ is the polytropic index. Making the transformation $\rho_m =
\rho_c\theta^n$, with $\rho_c$ being a constant and $\theta$ a
function of the dimensionless coordinate $\xi$ defined by $r =
[(n+1)K\rho_c^{1/n-1}/4\pi G]^{1/2}\xi$, the Euler equation for a
static fluid (\ref{qeuler}) becomes the Lane-Endem equation:
\begin{equation}
\frac{1}{\xi^2}\frac{\partial}{\partial\xi}\left(\xi^2
\frac{\partial\theta}{\partial\xi}\right)+\theta^n=0\;.
\end{equation}

The condensate density profile has a uniform behavior on the center of
the system, decreasing towards the border. Because of this behavior,
the quantum potential term contribution in the center is smaller than
the non-linear interaction term. On the other hand, on the border, the
contribution of the quantum potential is significant. When the number
of particles is large, the uniform region is increased and, in this
condition, we can obtain an analytical solution by neglecting the
quantum potential term in the equation (\ref{qeuler}).
This is the Thomas-Fermi approximation, in which the condensate is a
fluid whose density profile is limited to a region of the space. The
equation of state has the polytropic index $n = 1$, $K =
2\pi\hbar^2a/m^3$ and the Lane-Endem equation has an analytic solution
given by
\begin{equation}
\theta(\xi) = \frac{\sin(\xi)}{\xi}\;,
\end{equation}
with the appropriate boundary condition $\theta(0)=1$, which gives
$\rho_m(r=0) = \rho_c$. Thus $\rho_c$ is recognized as the central
density of the condensate. To calculate the radius $R$ of the
condensate, we impose the condition $\rho_m(R(\xi_0)) = 0$, which
gives $\xi_0 = \pi$ and:
\begin{equation}\label{radius0}
R = \pi \sqrt{\frac{\hbar^2 a}{Gm^3}}\;.
\end{equation}

In current theories for axions, these particles are assumed to
have no electric charge, but they can have a very small mass in the
range from $10^{-6}\; eV$ to $1\; eV$. In reference \cite{ran08} the authors
obtained an upper limit for the self-interacting dark matter
cross-section using results from X-ray, strong lensing, weak lensing,
and optical observation of the Bullet cluster 1E 0657-56. Based on
this cross-section, Harko et al. \cite{har12} estimated the upper
limit for the scattering lenght, $a<10^{-21}m\; (10^{-6}fm)$.
Using these data in (\ref{radius0}) we estimate the dark matter
condensate radius to lie in the range $10^{-2}\; pc$-$10^{7}\;
pc$. This radius range encompasses the size of dark matter halos in
typical galaxies, indicating that the axion Bose-Einstein condensate
is a viable candidate to represent dark matter halos in galaxies. As
we will see in the next sections, observational radii data constrain
the mass range even further.

\section{Spin-1 Particles Bose-Einstein Condensate}
\label{secspin1}

To derive an effective low energy Hamiltonian of a spin-1 condensate,
we introduce a spinor field operator $\hat{\Psi}_m({\bf r})$ (where $m
= -1,0,1$) corresponding to a field annihilation operator for a
dark matter particle in the spin state $|1,m\rangle$. The Hamiltonian
operator can be written in terms of these field operators as follows
\cite{ho98,fet71}
\begin{equation}
\hat{H}=\int d{\bf r}\sum_{m}\hat{\Psi}_m^\dagger({\bf r}) T_m({\bf
  r})\hat{\Psi}_m({\bf r})+\frac{1}{2}\iint d{\bf r}d{\bf
  r}'\sum_{m,m'}\hat{\Psi}_m^\dagger({\bf r})
\hat{\Psi}_{m'}^\dagger({\bf r}')V_{m,m'}({\bf r},{\bf
  r}')\hat{\Psi}_m({\bf r})\hat{\Psi}_{m'}({\bf r}')\;,
\end{equation}
where $T_m({\bf r})=K_m({\bf r})+U_m({\bf r})$ is the kinetic energy
plus the external potential energy of particles with spin $m$ and
$V_{m,m'}({\bf r},{\bf r}')$ is the interaction potential between
particles with spin $m$ and $m'$.

In reference \cite{ho98}, the author observed that the interactions
between particles are different for distinct spins. Then the system
symmetries lead the two particles vector state to be unchangeable
under the permutation of the particles. Therefore, one must have
$(-1)^{\cal F + J} = 1$ where $\cal F$ is the total spin of the system
and $\cal J$ is the angular momentum between the particles. In the low
energy dynamic of the system, we only consider the $s$-wave scattering
caracterized by two-body collisions with small momentum transfers and
represented by a $\delta$ function in coordinate space. Thus ${\cal J}
= 0$ and $\cal F$ must be even and range from 0 to $2f$, where $f$ is
the spin of the particles. For a system of $f=1$ bosons, we have
\begin{equation}
V_{m,m'}({\bf r},{\bf r}')=\delta({\bf r}-{\bf r}')\times [\lambda_0
  \delta_{m+m',0}+\lambda_2 (\delta_{m+m',2}+\delta_{m+m',-2})]\;,
\end{equation}
where $\lambda_{\cal F} = \frac{2\pi\hbar^2 a_{\cal F}}{m}$, with
$a_{\cal F}$ being the scattering length between particles of total
spin $\cal F$. Likewise, the relation ${\bf F}_1\cdot {\bf F}_2=
(\delta_{m+m',2}+\delta_{m+m',-2})-2\delta_{m+m',0}$, where ${\bf F}$
is the angular momentum operator, and the completeness relation $1=
\delta_{m+m',0} +\delta_{m+m',2}+\delta_{m+m',-2}$, we have the
general form of this interaction,
\begin{equation}
V_{m,m'}({\bf r},{\bf r}')=\delta({\bf r}-{\bf r}')[c_0+c_2 {\bf
    F}_1\cdot {\bf F}_2]\;,
\end{equation}
where $c_0 = \frac{\lambda_0+2\lambda_2}{3}$ and $c_2 =
\frac{\lambda_2-\lambda_0}{3}$.

Finally the effective low energy second quantized Hamiltonian is
\begin{eqnarray}
\hat{H} &=& \int d{\bf r} \sum_m\hat{\Psi}_m^\dagger({\bf r})T_m({\bf
  r}) \hat{\Psi}_m({\bf r})+\nonumber\\ &+&\frac{c_0}{2} \int d{\bf
  r}\sum_{m,m'}\hat{\Psi}_m^\dagger({\bf
  r})\hat{\Psi}_{m'}^\dagger({\bf r}) \hat{\Psi}_m({\bf r})
\hat{\Psi}_{m'}({\bf r})+\nonumber\\ &+&\frac{c_2}{2} \int d{\bf
  r}\sum_{m,m'}\hat{\Psi}_m^\dagger({\bf
  r})\hat{\Psi}_{m'}^\dagger({\bf r}){\bf F}_1\cdot {\bf F}_2
\hat{\Psi}_m({\bf r}) \hat{\Psi}_{m'}({\bf r})\; .
\label{hspin1}
\end{eqnarray}

As the interatomic interaction is elastic and conserves the total
number and total spin of particles, the hamiltonian is invariant under
the global $U(1)_\phi=\{e^{i\phi}1|\phi\in [0,2\pi]\}$ gauge
transformation, the $SO(3)_{\bf
  F}=\{e^{-iF_z\alpha}e^{-iF_y\beta}e^{-iF_x\gamma}\}$ rotation in
spin space where $(\alpha,\beta,\gamma)$ are the Euler angles, and
time reversal $\Theta=\{{\bf 1},{\cal T}\}$ where ${\cal T}$ is the
time-reversal operator.

At low temperature, there are phases regarding to the broken
symmetries of the ground state. To describe these phases, we break
the global gauge and the rotation in spin space symmetries
substituting the fields operator $\hat{\Psi}_m$ by the Bose condensate
$\Psi_m({\bf r})=\langle \hat{\Psi}_m ({\bf r})\rangle$ in the
Hamiltonian (\ref{hspin1}) obtaining the energy functional ${\cal H}
(\Psi_m)=\langle \hat{H}\rangle$, and construct the Lagrangian density
as ${\cal L} = -\frac{\hbar}{2i}\sum_{m}[(\Psi_m^*)(\partial_t\Psi_m)
  - (\partial_t\Psi_m^*)(\Psi_m)] - {\cal H}(\Psi_m)$.
Using the Euler-Lagrange equations, we obtain the Gross-Pitaevskii
equation for the spin-1 condensate,
\begin{eqnarray}
i\hbar\frac{\partial\Psi_m}{\partial t} =
\left[-\frac{\hbar^2\nabla^2}{2M} + U(\textbf{r})\right]\Psi_m +
c_0n({\bf r})\Psi_m + c_2\sum_{m'} \langle\textbf{F}\rangle\cdot
\textbf{f}_{mm'} \Psi_{m'}\;,
\label{sGPE}
\end{eqnarray}
where $n(\textbf{r}) = \langle\hat{n}\rangle = \sum_{m}
|\Psi(\textbf{r})_m|^2$ is the total condensate density, $
\langle\hat{\textbf{F}}\rangle = (F_x,F_y,F_z)$ is the average of the
angular momentum components, and $ \textbf{f}_{mm'}=\Psi_m{\bf F}\Psi^
*_{m'}$ is the angular momentum projection of state $\Psi_m$. For the
sake of simplicity we assume equal masses $M$ for different spin
particles and equal external potential, $U_m({\bf r})=U({\bf r})$.

Considering the total number of particles as being conserved, we can
write $\Psi_m(\textbf{r},t)=\psi_m({\bf r})e^{-i\mu t/\hbar}$ where
$\psi_m({\bf r})$ is a steady state and $\mu$ is the chemical potential,
and therefore we get the time-independent Gross-Pitaevskii equation
\begin{equation}
\mu\psi_m({\bf r}) = \left[-\frac{\hbar^2\nabla^2}{2M} +
  U(\textbf{r})\right]\psi_m({\bf r}) + c_0n({\bf r})\psi_m({\bf r}) +
c_2\sum_{m'} \langle{\bf F}\rangle\cdot \textbf{f}_{mm'}
\psi_{m'}({\bf r})\;.
\label{tigpes1}
\end{equation}

The equation (\ref{tigpes1}) had already been derived in \cite{ho98},
where it was solved by the use of the energy balance. In the present
work, we take a different approach and solve this equation by
considering that the phases are related to the spin rotational. In
order to identify these phases, let the condensate be rewritten by
$\psi_m({\bf r})=\sqrt{n({\bf r})}\zeta_m({\bf r})$, where ${\bf
  \zeta}=(\zeta_1,\zeta_0,\zeta_{-1})^T$ is a normalized spinor that
transform by the spin rotation
$D(\alpha,\beta,\gamma)=e^{-iF_z\alpha}e^{-iF_y\beta}e^{-iF_x\gamma}$. Kawaguchi
and Ueda showed in \cite{kaw11} that there are two possible phases for
the ground state related to the inert states that have continuous
isotropy groups. There are the ferromagnetic state where ${\bf
  \zeta}=(1,0,0)^T$ and the polar (antiferromagnetic) state where ${\bf
  \zeta}=(0,1,0)^T$ depends on the signal of $c_2$. The Physics of the
phase diagram is simple: $\langle {\bf F}\rangle$ is zero for $c_2>0$
(i.e. $\lambda_2>\lambda_0$) and the ground state is polar
(antiferromagnetic), or $\langle {\bf F}\rangle$ is maximal for
$c_2<0$ (i.e. $\lambda_2<\lambda_0$) and the ground state is
ferromagnetic. Hence, there are two distinct cases:

(I) Polar state, where the spinor $\zeta$ and the density $n({\bf
  r})$ in the ground state are:
\begin{equation}
\zeta=\frac{e^ {i\theta}}{\sqrt{2}}(e^
           {-i\alpha}\sin\beta,\cos\beta,e^ {i\alpha}\sin\beta)^T\;, 
\end{equation}
\begin{equation}\label{np}
n_p({\bf r})=\frac{1}{c_0}[\mu-U({\bf r})-W({\bf r})]\;,
\end{equation}
where $W({\bf r})=\frac{\hbar^ 2}{2M}\frac{\nabla^
  2\sqrt{n}}{\sqrt{n}}$ is the quantum potential.

(II) Ferromagnetic state, where the spinor $\zeta$ and the density
$n({\bf r})$ in the ground state are:
\begin{equation}
\zeta=e^ {i(\theta-\gamma)}(e^ {-i\alpha}\cos^
2\frac{\beta}{2},\sqrt{2}\cos\frac{\beta}{2}\sin\frac{\beta}{2},e^
{i\alpha}\sin^ 2\frac{\beta}{2})^T\;,
\end{equation}
\begin{equation}\label{nf}
n_f({\bf r})=\frac{1}{\lambda_2}[\mu-U({\bf r})-W({\bf r})]\;.
\end{equation}

Considering that the external potential obeys the Poisson equation and
the gas is composed of a large number of particles, we perform the
Thomas-Fermi approximation neglecting the quantum potential in
equations (\ref{np}) and (\ref{nf}). Thus, we have
\begin{equation}
\nabla^2n_p({\bf r})=-\frac{4\pi G M}{c_0}n_p({\bf r})\;,
\end{equation}
\begin{equation}
\nabla^2n_f({\bf r})=-\frac{4\pi G M}{\lambda_2}n_f({\bf r})\;,
\end{equation}
which constitute the polytropic gas with index $n=1$ and the
coefficient $K=2c_0$ for polar state and $K=2\lambda_2$ for
ferromagnetic state. Using the parametrized analytic solution of
Lane-Endem equation as density profile, we can fit the condensate
radius $R$ using the interaction parameters, $a_0$ and $a_2$, and its
mass.  For the polar phase we have
\begin{equation}\label{rpolar}
 R_p = \pi \sqrt{\frac{\hbar^2 (a^p_0+2a^p_2)}{3GM^3}}
\end{equation}
and for the ferromagnetic phases
\begin{equation}\label{rferr}
 R_f = \pi \sqrt{\frac{\hbar^2 a^f_2}{GM^3}}\;. 
\end{equation}

If the dark matter condensate in the galaxy is in the ferromagnetic
phase, i. e. $a_0^f >a_2^f$, the halo radius depends only on the
scattering length $a_2^ f$ and the treatment is similar to the spin-0
model. However, if the dark matter condensate in the galaxy is in the
polar phase, i. e. $a_0^p <a_2^ p$, the halo radius depends on the
scattering lengths $a_0^ p$ and $a_2^p$. In this phase, there are two
parameters to be determined and using astronomical data we can
constrain the mass and scattering lengths of these particles, making
use of a Maximum Likelihood analysis, as shown in the next section.

\section{Statistical Analysis}
\label{secstatistical}

In order to constrain the mass of the dark matter particle and its
relevant scattering lengths, we use here a subset of the astronomical
data presented in \cite{auger}. This subset presents information on 42
dark matter dominated spiral galaxies ($f_{DM}\ge 0.5$, using Chabrier
initial mass function), studied by the weak lensing method. We are
interested in these galaxies' halo radii, which range from $1.39\;
kpc$ to $20.09\; kpc$.

Using these previously mentioned data, we can construct the
Likelihood function $\cal{L}$ for the halo radius $R(a)$ trough:
\begin{equation}
\label{Like}
{\cal{L}} \propto
\prod_{i=1}^{N} \exp
\left\{  - \frac{1}{2 \, \sigma^2_i}
\left[
R(a) - r_{i}
\right]^2
\right\} \; ,
\end{equation}
\noindent where $R(a)$ represents the theoretical radius obtained from
eqs. (\ref{radius0}), (\ref{rpolar}) or (\ref{rferr}), $r_{i}$ are the
data taken from observations and $\sigma_{i}$ are the errors
associated with these measurements. The errors were not available in
the mentioned work, hence we have decided to overestimate these
quantities and assume a worst-case scenario in which they would have
half the value of each radius measurement.

For the particle mass, for both the spin-0 and spin-1 cases, we have
chosen at first to use the lower bound of the mass range for the
axion, $10^{-6}\; eV$ (see \cite{sikivie1} and references therein).

\begin{figure}[!htp]
\begin{center}

\includegraphics[scale=0.69]{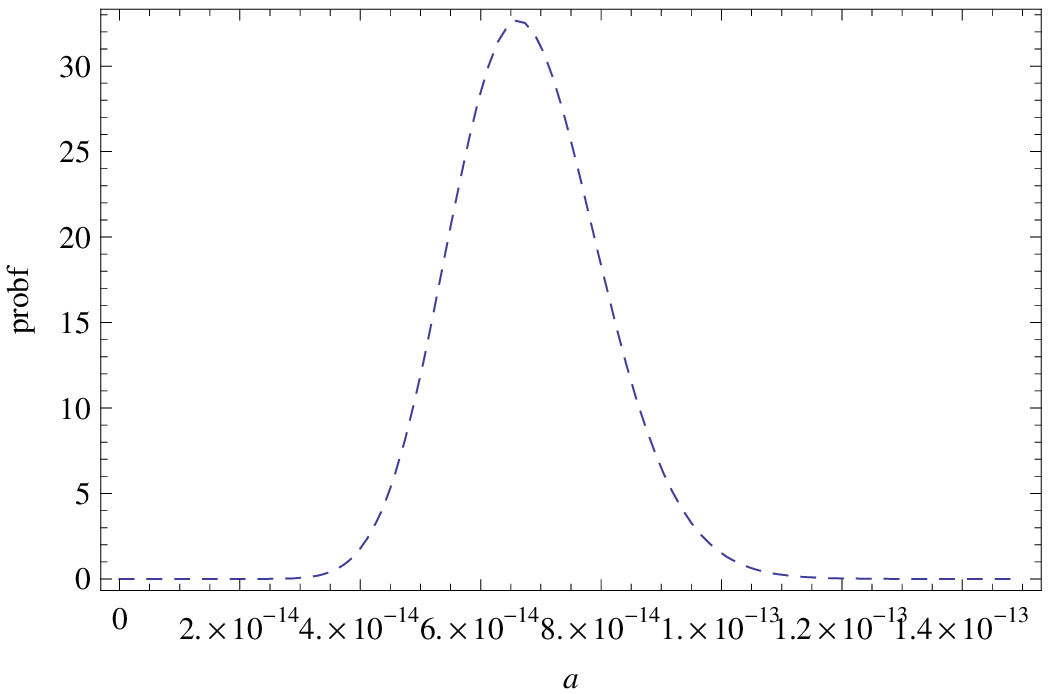}
\hspace{0.3cm}
\includegraphics[scale=0.71]{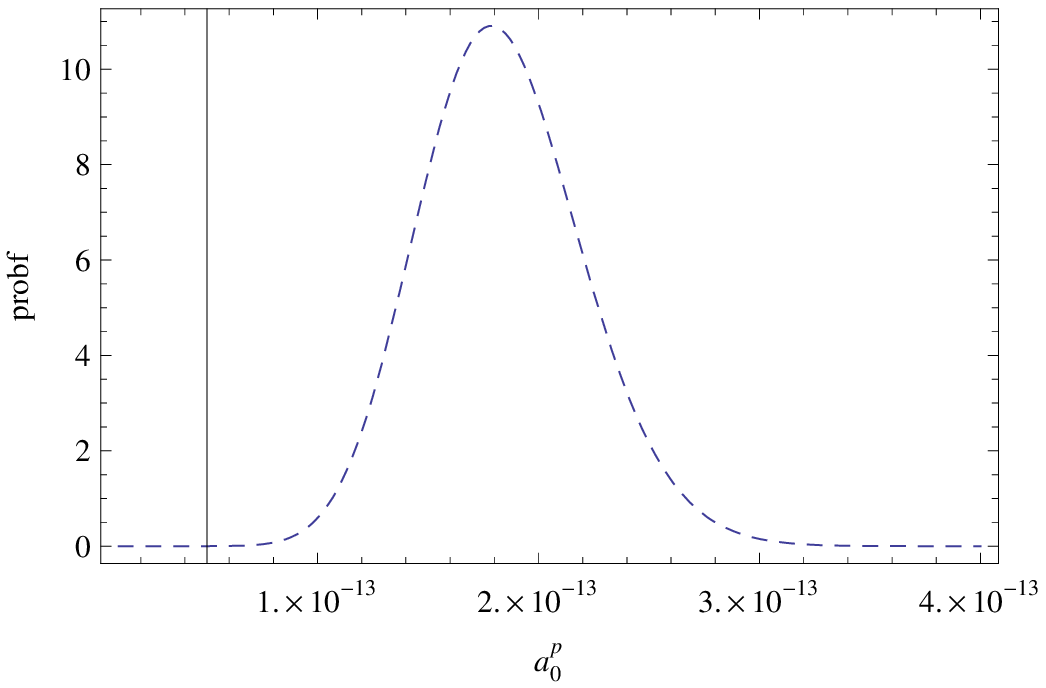}

\caption{Probability density function for the scattering length (in
  $fm$) using 42 spiral galaxies radii and a dark matter particle mass of
  $10^{-6}\; eV$. The left panel shows the spin-0 case, and the right
  panel shows the spin-1 case.}
\label{pdf1}

\end{center}
\end{figure}

For the spin-0 case, in which there is only one scattering length $a$
to be determined, the probability density function constructed is
shown in the left panel of figure (\ref{pdf1}).

\vspace{0.5cm}

The peak of the probability density function constructed from the
Likelihood gives us the best fit for the parameter in
investigation. We can see in figure (\ref{pdf1}) that the best value
is around $10^{-14}\; fm$. Since we have overestimated the errors used
in calculations, the width of this curve is also overestimated.
 
For the spin-1 case, we have to make some additional assumptions. We
assume that the ferromagnetic phase scattering length $a_{2}^{f}$ and
the polar phase scattering length $a_2^{p}$ have the same value, and
that this does not differ from $a$ obtained for the spin-0 case. Using
the previously calculated value as an input, we find the probability
function for the scattering length $a_0^{p}$ as shown in the right
panel of fig. (\ref{pdf1}). The best fit value for this quantity is
around $10^{-13}\; fm$.

\begin{figure}[!htp]
\begin{center}

\includegraphics[scale=0.69]{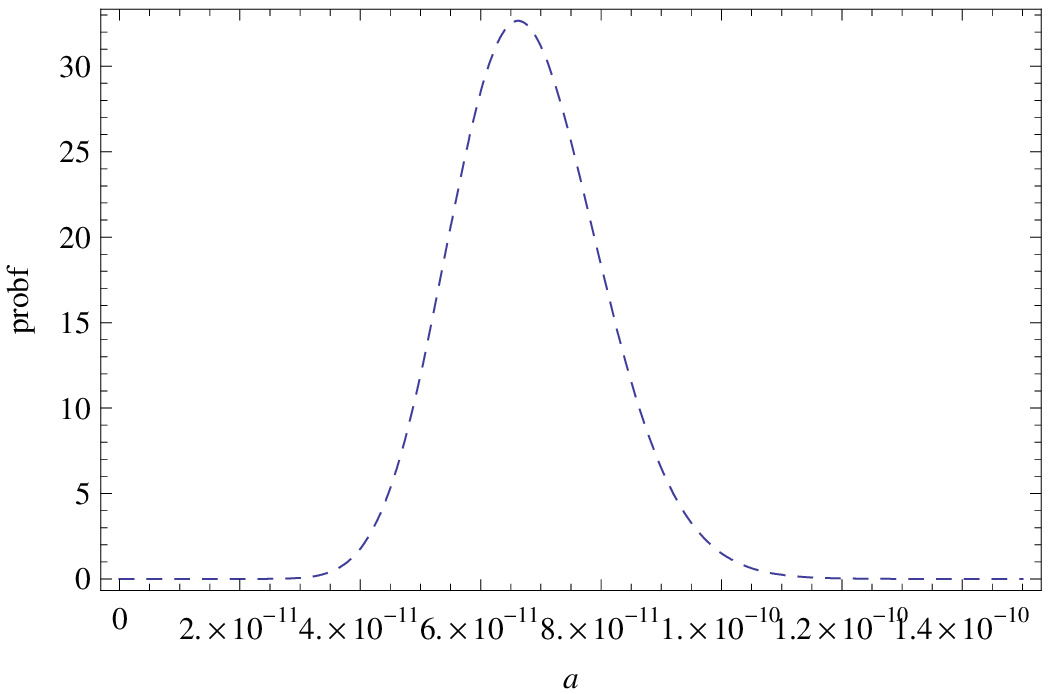}
\hspace{0.3cm}
\includegraphics[scale=0.71]{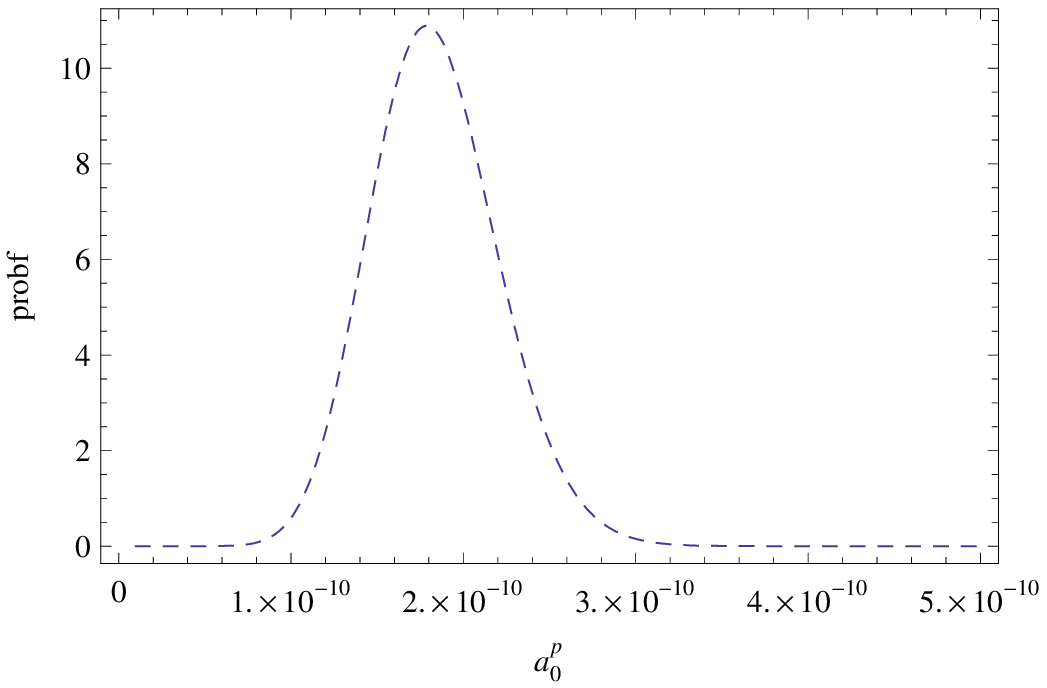}

\caption{Probability density function for the scattering length
  (in $fm$) using 42 spiral galaxies radii and a dark matter particle mass of
  $10^{-5}\; eV$. The left panel shows the spin-0 case, and the right
  panel shows the spin-1 case.}
\label{pdf2}

\end{center}
\end{figure}

We also performed the same analysis using slightly higher values for
the mass. Fig. (\ref{pdf2}) shows the case for a dark matter particle
with a mass of the order of $10^{-5}\; eV$. The scattering length in
this case is found to be $\sim 10^{-11}\;(10^{-10})\; fm$ for the
spin-0 (spin-1) case.

\vspace{0.5cm}

\begin{figure}[!htp]
\begin{center}

\includegraphics[scale=0.69]{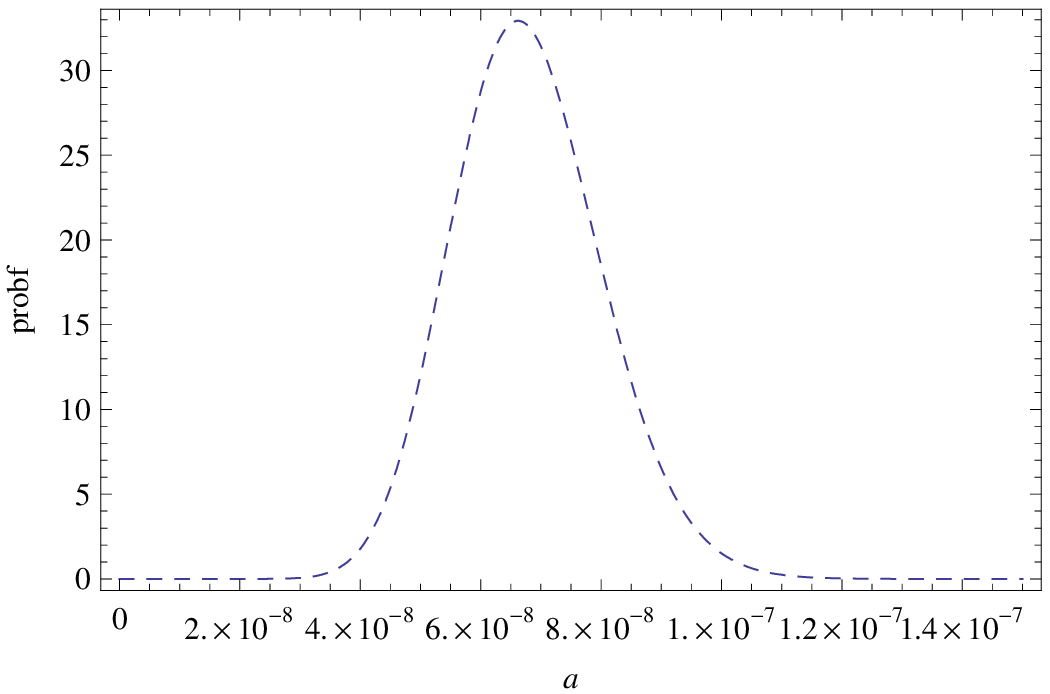}
\hspace{0.3cm}
\includegraphics[scale=0.71]{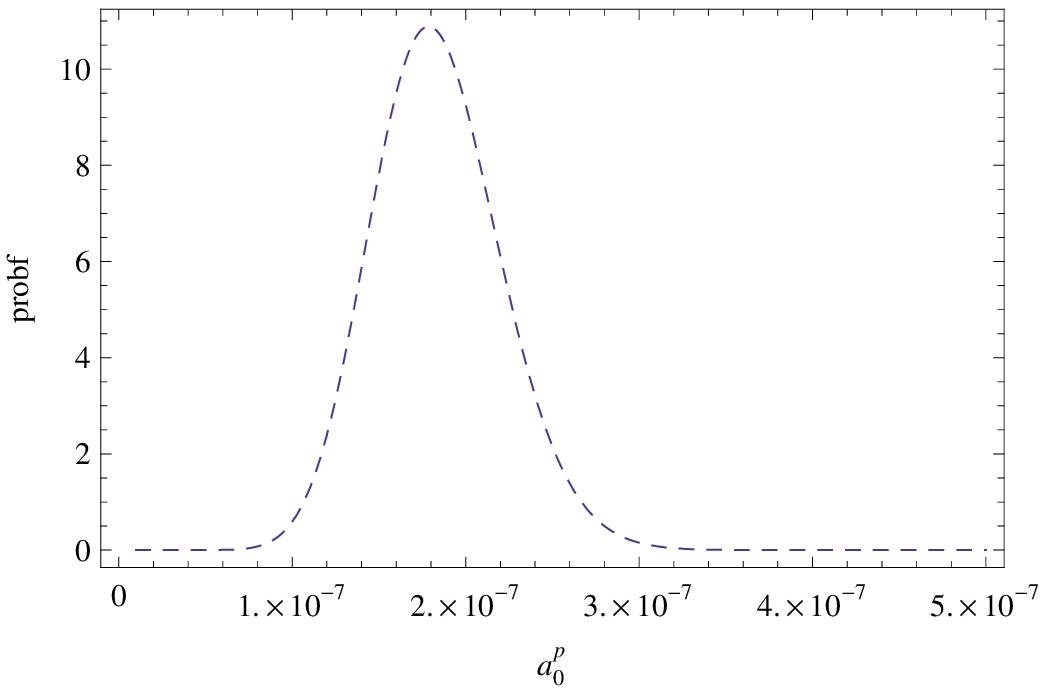}

\caption{Probability density function for the scattering length (in
  $fm$) using 42 spiral galaxies radii and a dark matter particle mass of
  $10^{-4}\; eV$. The left panel shows the spin-0 case, and the right
  panel shows the spin-1 case.}
\label{pdf3}

\end{center}
\end{figure}

The case of a particle with a mass of the order $10^{-4}\; eV$ is
shown in Fig. (\ref{pdf3}). The scattering length in this case is
found to be $\sim 10^{-8}\;(10^{-7})\; fm$ for the spin-0 (spin-1)
case.

Considering larger masses causes the scattering length to lie outside
the upper bound referred to in section \ref{secspin1}.

\vspace{0.5cm}

\begin{figure}[!htp]
\begin{center}

\includegraphics[scale=0.69]{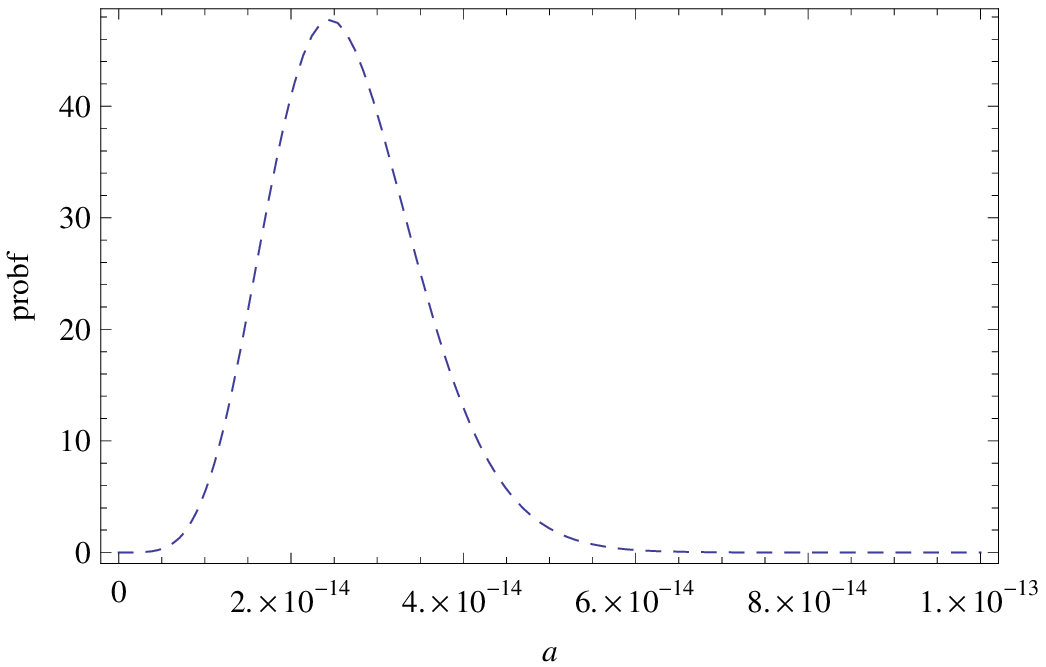}
\hspace{0.3cm}
\includegraphics[scale=0.71]{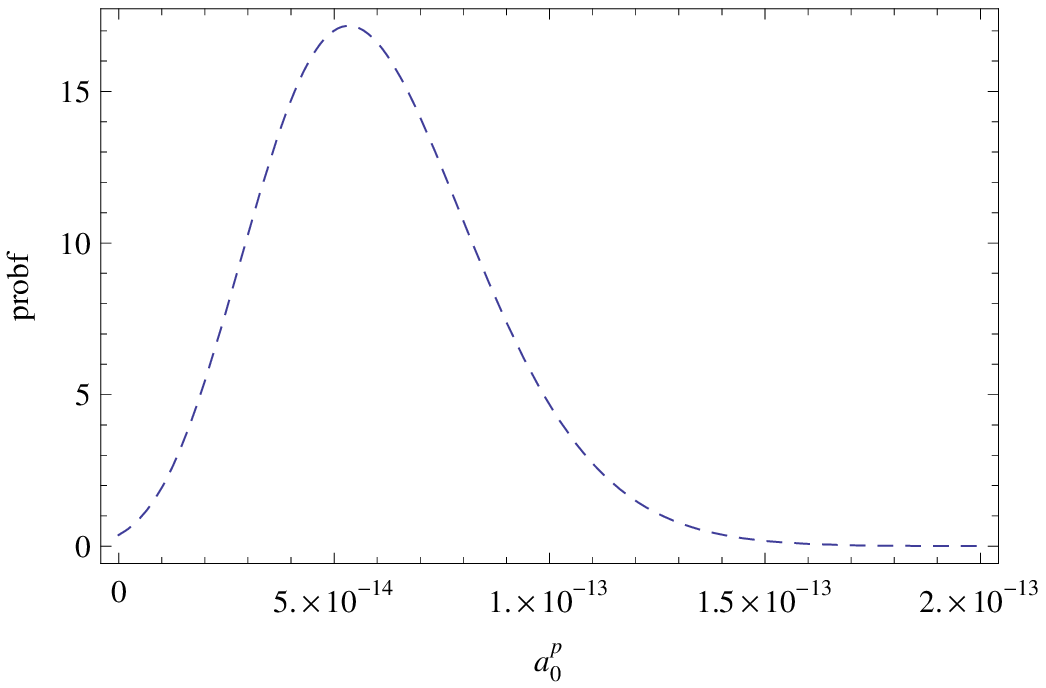}

\caption{Probability density function for the scattering length (in
  $fm$) using 19 LSB galaxies radii and a dark matter particle mass of $10^{-6}\;
  eV$. The left panel shows the spin-0 case, and the right panel shows
  the spin-1 case.}
\label{lsb6}

\end{center}
\end{figure}

\begin{figure}[!htp]
\begin{center}

\includegraphics[scale=0.69]{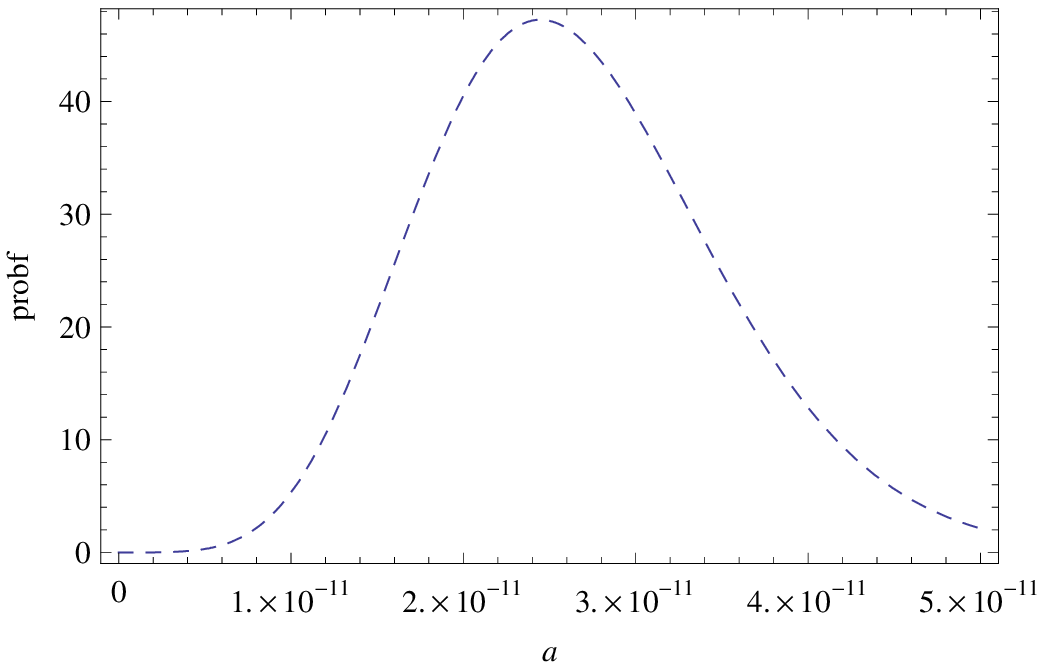}
\hspace{0.3cm}
\includegraphics[scale=0.71]{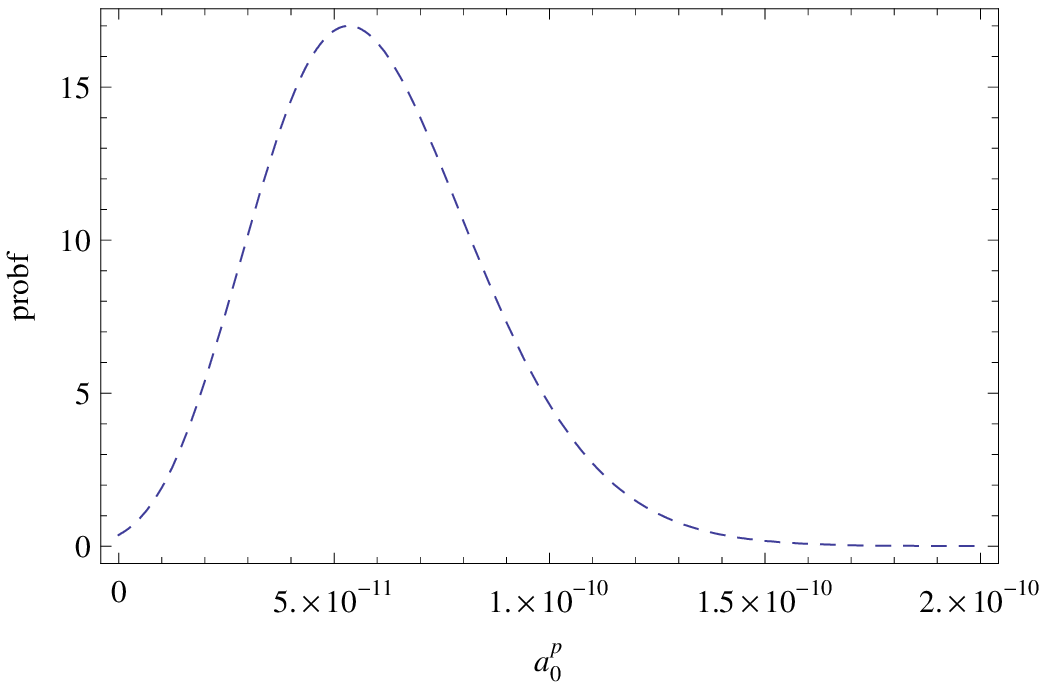}

\caption{Probability density function for the scattering length (in
  $fm$) using 19 LSB galaxies radii and a dark matter particle mass of $10^{-5}\;
  eV$. The left panel shows the spin-0 case, and the right panel shows
  the spin-1 case.}
\label{lsb5}

\end{center}
\end{figure}

\begin{figure}[!htp]
\begin{center}

\includegraphics[scale=0.69]{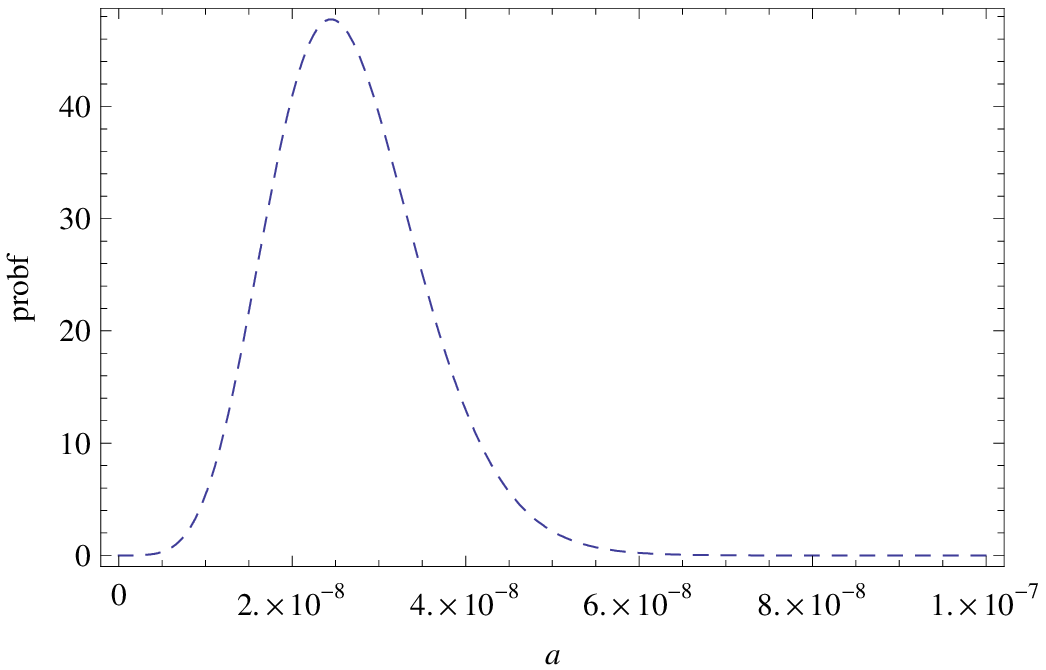}
\hspace{0.3cm}
\includegraphics[scale=0.71]{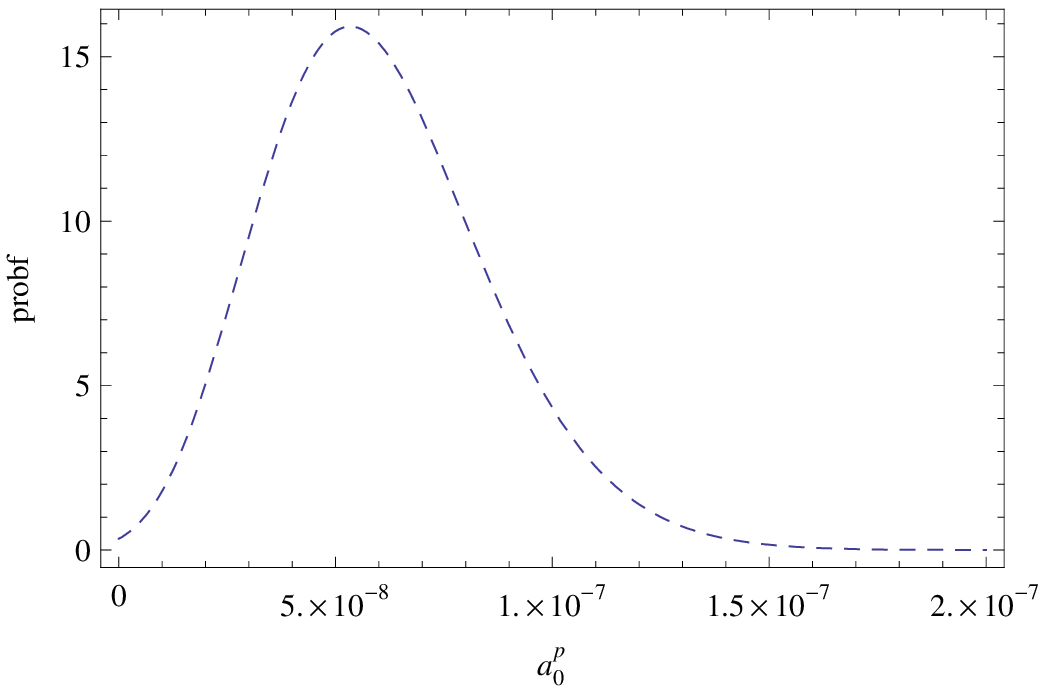}

\caption{Probability density function for the scattering length (in
  $fm$) using 19 LSB galaxies radii and a dark matter particle mass of $10^{-4}\;
  eV$. The left panel shows the spin-0 case, and the right panel shows
  the spin-1 case.}
\label{lsb4}

\end{center}
\end{figure}

Another sample, now containing 19 low surface brightness (LSB, dark
matter dominated objects with $f_{DM}>0.9$) galaxies radii ranging
from 1.2 to 19.6 $kpc$, obtained from \cite{lsb}, have been analysed
using the same method. The corresponding plots are shown in
Figs. \ref{lsb6}-\ref{lsb4}.

\vspace{0.5cm}

\vspace{0.5cm}

\vspace{0.5cm}

The values of the scattering lengths obtained with the use of LSB
galaxies do not differ from the previous results. A possible
explanation for this behavior is that our expressions for the radius
of the BEC fluid derived from Thomas-Fermi approximation depend only
on the dark matter particle mass, and not on the astrophysical objects
masses. Since the radii ranges are similar in both the galaxies sets
we used, it is not surprising that the scattering lengths are also
similar.

\section{Conclusions}
\label{secconclusion}

In this work, we have been able to develop a theory for a spin-1
Bose-Einstein condensate composed of WISPs in terms of their
scattering length and their masses. We could identify two phases of
this condensation state, namely, a polar phase where the total angular
momentum is zero, and a ferromagnetic phase where the total momentum
is in the maximal projection.

Considering also spin-0 particles (related to axions or axion-like
particles), it was possible to use this condensate to model dark
matter halos in galaxies and obtain their radii in terms of the same
parameters.

In order to constrain the values of scattering lengths and masses for
the two possible condensate phases, we proceeded to a statistical
analysis using a set of 42 dark matter dominated spiral galaxies and
19 LSB galaxies radii. Even though the set used presents a relativily
small number of data, we have been able to limit the mass of the
proposed dark matter particle to the range $10^{-6}-10^{-4}\; eV$ and
to find the lower bound for the scattering length to be $10^{-14}\;
fm$. We recall here that the upper bound for this quantity have been
estimated to be at the value $10^{-6}\; fm$ by the use of colliding
clusters data in \cite{har12}.

Using this method we were not able to determine the dark matter
particle's spin, because of the similarity between the spin-0 and
ferromagnetic phase expressions for the galaxy radius as a function of
the scattering length.  Even considering the fluid to be composed of
spin-1 particles, we were not able to distinguish the corresponding
two phases by the use of this statistical analysis, since they are
linked by the use of $a_{2}^{f}$ as an input for $a_{2}^{p}$in the
construction of the probability functions . A more comprehensive
analysis involving a significantly larger set of galaxy data may be
necessary to establish this distinction beyond doubt. Nevertheless,
one of the objectives of this paper is to show that this method can
bring important information on the characteristics of the dark matter
fluid.

The study of the condensate excitations and speed of sound could also
allow the determination of the spin state of the fluid's
particles. These features can be related to the ocurrence of
observable caustics and cusps in the galactic halo phase space
\cite{sikivie2} and will be the subject of future work.

We would like to point out that one interesting feature of the present
approach is the use of macroscopic quantities (e. g., astronomically
determined galaxies' radii) to gather information on microscopic
parameters (particle masses and scattering lengths) related to quantum
aspects of the nonrelativistic fluid. This correspondence is possible
because we are treating a Bose-Einstein condensate. Another quantum
fluid where this relation can be made is the degenerated Fermi fluid
with half integer spin particles. This system will be treated in an
upcoming work.

\begin{acknowledgments}
The authors would like to thank Andr\'e M. Lima for help in the early
phases of the present work. M. O. C. P. is grateful to C. F. Martins
for presenting this subject. J. C. C. S. thanks CAPES (Coordena\c
c\~ao de Aperfei\c coamento de Pessoal de N\'\i vel Superior) for
financial support.
\end{acknowledgments}


\begin{thebibliography}{99}

\bibitem{wmap} E. Komatsu et al., {\it Astrophys. J.}, {\bf 192} 18
  (2011)

\bibitem{beringer} J. Beringer et al. (Particle Data Group), {\it
  Phys. Rev. D} {\bf 86} 010001 (2012)


\bibitem{jaeckel} P. Arias et al., {\it J. Cosmol. Astropart. Phys.}
  {\bf 06} 013 (2012)

\bibitem{nel11} A. E. Nelson and J. Scholtz, {\it Phys. Rev. D} {\bf
  84}, 103501 (2011)


\bibitem{peccei} R. D. Peccei, and H. R. Quinn, {\it Phys. Rev. Lett.}
  {\bf 38} 1440 (1977); R. D. Peccei, and H. R. Quinn, {\it
    Phys. Rev. D} {\bf 16} 1791 (1977)

\bibitem{feng10} J. Feng, {\it Ann. Rev. Astron. Astrophys.} {\bf 48}
  495 (2010)

\bibitem{axion01} S. -J. Sin, {\it Phys. Rev. D} {\bf 50}, 3650
  (1994); W. Hu, R. Barkana, and A. Gruzinov, {\it Phys. Rev. Lett.}
  {\bf 85}, 1158 (2000); J. A. V\'elez P\'erez, {\it Phys. Lett. B} {\bf
    671}, 174 (2009)

\bibitem{arias} P. Arias, J. Jaeckel, J. Redondo and A. Ringwald, {\it
  Phys. Rev. D} {\bf 82} 15018 (2010)

\bibitem{baker} O. Baker et al., {\it Phys. Rev. D} {\bf 85} 035018
  (2012)

\bibitem{sikivie1} P. Sikivie and Q. Yang, {\it Phys. Rev. Lett.}
  {\bf 103}, 111301 (2009)


\bibitem{boh07} C. G. Böhmer and T. Harko, {\it
  J. Cosmol. Astropart. Phys.} {\bf 06}, 025 (2007)

\bibitem{chavanis} P.-H. Chavanis, {\it Phys. Rev. D} {\bf 84} 043531
  (2011)
\bibitem{chavanis2} P.-H. Chavanis and L. Delfini, {\it Phys. Rev. D} {\bf 84} 043532
  (2011)

\bibitem{lora} V. Lora et al., {\it J. Cosmol. Astropart. Phys.}
  {\bf 02} 011 (2012)

\bibitem{robles} V. H. Robles and T. Matos, {\it
  Mon. Not. Roy. Astron. Soc.} {\bf 422}, 282 (2012)

\bibitem{stenger98} J. Stenger et al., {\it Nature} {\bf 396}, 345
  (1998)

\bibitem{pethick} C. J. Pethick and H. Smith, {\it Bose-Einstein
  Condensation in Dilute Gases} (2001) Cambridge-Press


\bibitem{ran08} S. W. Randall, M. Markevitch, D. Clowe,
  A. H. Gonzalez, and M. Bradac, {\it Astrophys. J.} {\bf 679}, 1173
  (2008)

\bibitem{har12} T. Harko, and G. Mocanu, {\it Phys. Rev. D} {\bf 85}, 084012 (2012)



\bibitem{ho98} T.-L. Ho, {\it Phys. Rev. Lett.} {\bf 81}, 742 (1998)
 
\bibitem{fet71} A. Fetter and J. D. Walecka, {\it Quantum Theory of
  Many-Particle System} (1971) Dover

\bibitem{kaw11} Y. Kawaguchi and M. Ueda, {\it Phys. Rev. A} {\bf 84},
  053615 (2011)

\bibitem{auger} M. W. Auger et al., {\it Astrophys. J.} {\bf 724}, 511
  (2010)

\bibitem{lsb} C. Trachternach et al., {\it Astron. J.} {\bf 136}, 2720 (2008)

\bibitem{sikivie2} L. D. Duffy and P. Sikivie, {\it Phys. Rev. D} {\bf
  78}, 063508 (2008)

\end{thebibliography}
\end{document}